\documentclass[10pt,conference]{IEEEtran}
\usepackage[utf8]{inputenc}
\usepackage[noadjust]{cite}
\usepackage{graphicx}
\usepackage{afterpage}
\usepackage[caption=false]{subfig}
\usepackage{multirow}
\usepackage{multicol}
\usepackage{array}
\usepackage{xcolor}
\usepackage{float}
\usepackage[normalem]{ulem}
\usepackage{listings}
\usepackage{makecell}
\usepackage[hyphens]{url}
\usepackage{hyperref}
\usepackage{ulem}
\usepackage{xspace}
\usepackage{amsmath}
\usepackage{algorithm}
\usepackage{algorithmic}
\usepackage{amssymb}
\usepackage{censor}

\newcommand{\sng}{SuperMUC-NG\xspace}
\newcommand{\cmuc}{CooLMUC-3\xspace}
\newcommand{\lrz}{LRZ\xspace}
\newcommand{\ethz}{ETH\xspace}

\begin{document}

\title{Correlation-wise Smoothing: Lightweight Knowledge Extraction for HPC Monitoring Data}
\author{\IEEEauthorblockN{Alessio Netti\IEEEauthorrefmark{1}\IEEEauthorrefmark{2}, Daniele Tafani\IEEEauthorrefmark{3}, Michael Ott\IEEEauthorrefmark{1} and Martin Schulz\IEEEauthorrefmark{1}\IEEEauthorrefmark{2}}
\IEEEauthorblockA{\IEEEauthorrefmark{1}Leibniz Supercomputing Centre, Garching bei M\"unchen, Germany \\ Email: \texttt{\{alessio.netti,  michael.ott\}@lrz.de}}
\IEEEauthorblockA{\IEEEauthorrefmark{2}Technical University of Munich, Garching bei M\"unchen, Germany \\ Email: \texttt{schulzm@in.tum.de}}
\IEEEauthorblockA{\IEEEauthorrefmark{3}Fujitsu Enabling Software Technology GmbH, M\"unchen, Germany \\ Email: \texttt{daniele.tafani@fujitsu.com}}}

\maketitle

\begin{abstract}
Modern \emph{High-Performance Computing} (HPC) and data center operators rely more and more on data analytics techniques to improve the efficiency and reliability of their operations. They employ models that ingest time-series monitoring sensor data and transform it into actionable knowledge for system tuning: a process known as \emph{Operational Data Analytics} (ODA). However, monitoring data has a high dimensionality, is hardware-dependent and difficult to interpret. This, coupled with the strict requirements of ODA, makes most traditional data mining methods impractical and in turn renders this type of data cumbersome to process. Most current ODA solutions use ad-hoc processing methods that are not generic, are sensible to the sensors' features and are not fit for visualization.

In this paper we propose a novel method, called \emph{Correlation-wise Smoothing} (CS), to extract descriptive signatures from time-series monitoring data in a generic and lightweight way. Our CS method exploits correlations between data dimensions to form groups and produces image-like signatures that can be easily manipulated, visualized and compared. We evaluate the CS method on HPC-ODA, a collection of datasets that we release with this work, and show that it leads to the same performance as most state-of-the-art methods while producing signatures that are up to ten times smaller and up to ten times faster, while gaining visualizability, portability across systems and clear scaling properties.
\end{abstract}


\section{Introduction}
\label{section:introduction}

Modern compute facilities, be it in the form of \emph{High-Performance Computing} (HPC) or commercial data centers, are large and complex installations whose design, construction and operation pose significant challenges: they consume large amounts of energy and hence achieving energy efficiency is key~\cite{villa2014scaling}; further, as data center and HPC systems contain a large number of individual components, reliable operation is far from trivial~\cite{cappello2014toward}. In order to address such challenges, more and more compute centers adopt data analytics techniques to proactively analyze and improve system operations. These leverage the vast amounts of data produced by monitoring frameworks, which collect time-series sensor data from a large number of instrumented physical and software components at fine time scales, producing up to several millions of data points per second for large-scale installations~\cite{netti2019dcdb}. This data is then processed to extract knowledge about system behavior, which can ultimately be used to drive feedback control loops by tuning system knobs: approaches of this kind, which belong to the wider domain of multi-dimensional time-series analysis, are known as \emph{Operational Data Analytics} (ODA)~\cite{Bourassa:2019:ODA:3339186.3339210, bautista2019collecting}. 

\begin{figure}[b]
\centering
\includegraphics[width=0.45\textwidth,trim={0 0 0 0}, clip=true]{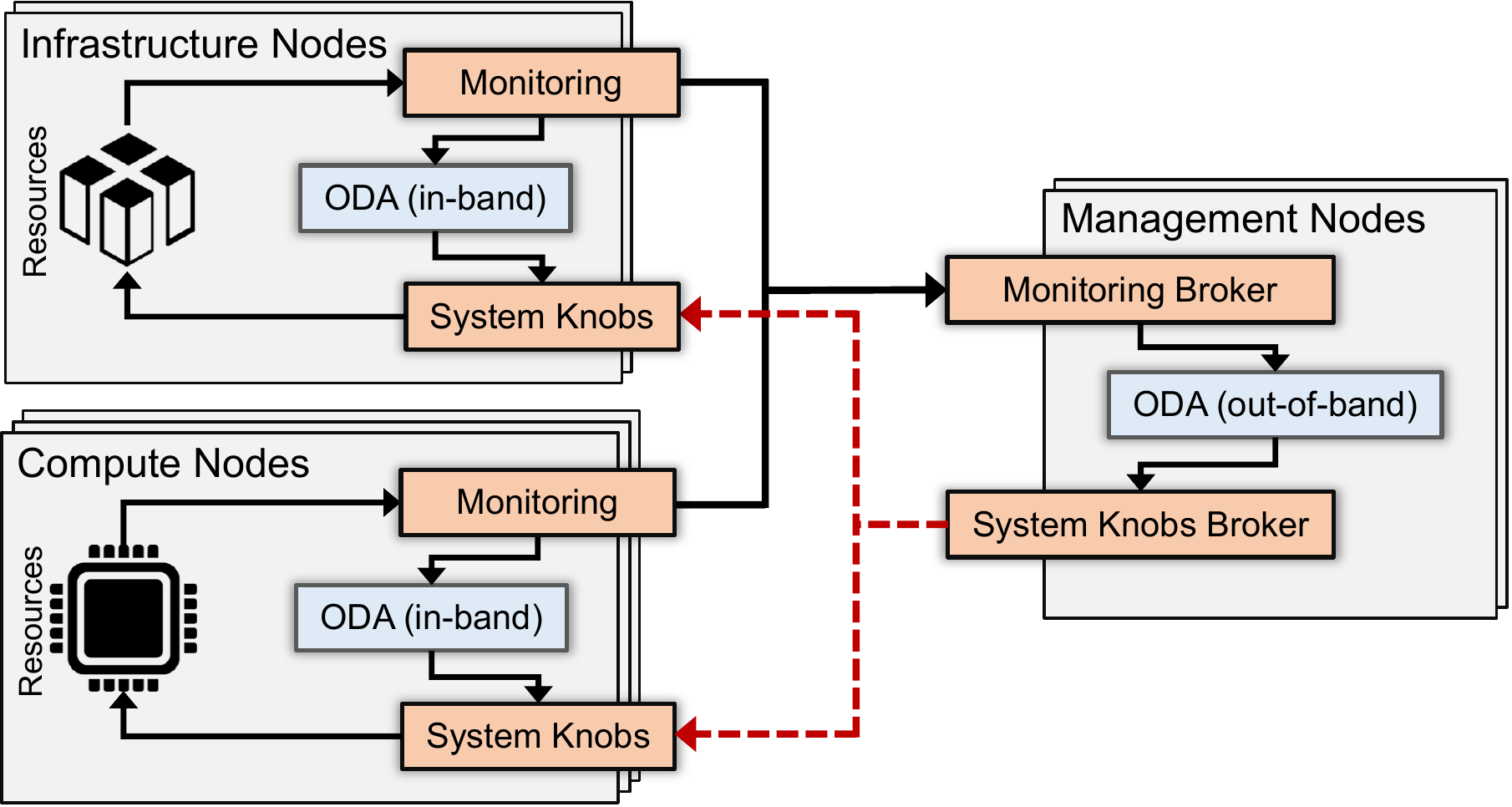}
\caption{Typical flow of data in a distributed system using ODA.}
\label{intro:odaexample}
\end{figure}

The majority of ODA techniques operate \emph{online} and follow the same framework of operation, shown in Figure~\ref{intro:odaexample}: \emph{monitoring} collects data from sensors of interest, which is then processed by ODA to produce a compact representation, i.e., a \emph{signature}. This is then fed to a model that is able to derive actionable knowledge, usually in the form of a new system setting. The latter is finally applied via a \emph{system knob}, which is a control interface for a certain component. Data center components offer a wide variety of knobs, such as CPU frequencies, fan speeds and water temperatures, up to high-level infrastructure settings~\cite{netti2019wintermute}. As such, ODA can be placed at any level of a data center with different requirements in terms of data sources, time scales, resource footprints and modes of operation: an ODA algorithm tuning CPU frequencies with the aim of optimizing energy efficiency~\cite{ozer2019towards} will run in compute nodes using local monitoring data (\emph{in-band}) and will act at very fine time scales (typically milliseconds), with strict requirements in terms of resource footprint and overhead on applications~\cite{ferreira2008characterizing}. On the other hand, an ODA algorithm optimizing a cooling system's inlet water temperature~\cite{conficoni2015energy} will operate in a dedicated management node (\emph{out-of-band}) at a coarse scale (typically minutes), but will need to cope with the large volumes of sensor data from the full breadth of the system. Due to these constraints, many traditional data mining methods~\cite{maimon2005data} are not applicable to ODA.

Many ODA algorithms use models belonging to the fields of traditional data mining and statistics~\cite{cohen2005capturing, hui2018comprehensive, lan2009toward, laguna2013automatic}, but machine learning-based ODA is increasingly gaining traction~\cite{tuncer2018online, bodik2010fingerprinting, borghesi2019semisupervised}. However, despite the existence of many effective ODA algorithms, there is a severe lack of generic methods to transform raw sensor data into signatures: monitoring data is inherently high-dimensional, architecture-dependent and hard to interpret, even for domain experts~\cite{gimenez2017scrubjay, ahlgren2018large}, leading to solutions that are limited to individual ODA problems, but lack any kind of generality. The latter aspect is critical for the development of \emph{exascale} HPC platforms, as their sustainable operation will require the coordinated use of ODA at all system levels and in turn a vast set of competences to train, operate and maintain the underlying models --- approaches to enable the sharing of knowledge and models across institutions are therefore necessary to simplify this daunting task. Looking particularly to signature algorithms tasked with processing sensor data, we observe the following requirements:

\begin{itemize}
    \item \textbf{Scalability}: the algorithm must be able to cope with large volumes of sensor data, stemming from large time spans and high dimensionality.
    \item \textbf{Footprint}: the algorithm's resource footprint must be minimal to not interfere with applications running on the system, which is critical for in-band ODA use cases.
    \item \textbf{Compression}: signatures must be significantly more compact than the original sensor data to simplify their use in training of models as well as inference.
    \item \textbf{Performance}: signatures must lead to ODA performance (e.g., classification accuracy) that is acceptable and predictable, to simplify the maintenance of models.
    \item \textbf{Visualizability}: signatures must be easily visualizable and interpretable, to enable intervention of human operators in the ODA processes.
    \item \textbf{Portability}: signatures must be generic and not system or architecture-dependent, as well as robust against the addition or removal of sensors over time.
\end{itemize}

\subsection{Related Work}
\label{section:relatedwork}

The problem of transforming data center monitoring data into actionable knowledge for ODA purposes has been dealt with in different ways. Tuncer et al.~\cite{tuncer2018online} propose a method that relies on computing statistical indicators (e.g., mean or standard deviation) of recent data from each available sensor, which are then used to build signatures. Bodik et al.~\cite{bodik2010fingerprinting} propose a similar approach, focusing instead on characterizing the distribution of each sensor's data by computing percentiles. Lan et al.~\cite{lan2009toward} use the raw time-series data of each sensor, concatenated to form a signature, thus preserving time information. Laguna et al.~\cite{laguna2013automatic} use the pairwise correlation matrix associated with the set of sensors as a signature, while Cohen et al.~\cite{cohen2005capturing} use an ensemble of statistical models to characterize the correlation of each sensor with a reference metric (e.g., a health indicator), from which the signature coefficients are extracted. Hui et al.~\cite{hui2018comprehensive} apply \emph{Singular Value Decomposition} to sensor data and compute a single entropy coefficient to characterize system state transitions. Finally, Borghesi et al.~\cite{borghesi2019semisupervised} use an autoencoder-based system to compress and subsequently extract sensor data.

Many classical techniques belonging to the wider domain of dimensionality reduction have been applied to monitoring data with varying degrees of success. \emph{Principal Component Analysis} (PCA)~\cite{abdi2010principal} and \emph{Independent Component Analysis}~\cite{hyvarinen2000independent} compress a multi-dimensional dataset to a lower-dimensionality space in which each dimension is a linear combination of the original ones, with higher weight associated to those that contribute more to the total variance. Guan et al.~\cite{guan2013adaptive} use a variant of this approach, named \emph{Most Relevant Component Analysis}, which aims at preserving some of the original dimensions intact. Techniques like \emph{Multidimensional Scaling}~\cite{cox2008multidimensional} or \emph{Vector Quantization}~\cite{gray1984vector} are also often employed. However, many traditional dimensionality reduction techniques have been proven to not work well in HPC and data center-specific ODA problems, such as fault detection, in which critical status indicators are not found in the metrics that contribute to most of the variance~\cite{tuncer2018online}. While many other approaches to process multi-dimensional time-series data are available, their fitness for system monitoring data is not clear. Among these we find \emph{Symbolic Aggregate Approximation}~\cite{lin2007experiencing} and \emph{Trend-value Approximation}~\cite{esmael2012multivariate}, which aggregate time-series data both on the time and value axes, and the \emph{Matrix Profile}~\cite{yeh2017matrix}, which computes data structures to aid in the identification of time-series motifs. Techniques belonging to the domain of spectral signal analysis~\cite{chan1999efficient} are also often used. 

In general, some of the methods discussed here exhibit very good performance in specific ODA use cases, but are not designed around generality, visualizability and scalability for large datasets with hundreds of dimensions. Moreover, they are extremely sensible to sensor properties such as scaling and order: this, coupled with the inability to manipulate the signatures in a simple way, hinders portability of models across systems and causes even minor changes in the monitoring infrastructure to be detrimental.

\begin{table*}[ht!]
\caption{An overview of the features of each segment in the HPC-ODA dataset collection.}
\label{table:hpcoda}
\begin{center}
\renewcommand{\arraystretch}{1.2}
\begin{tabular}{ccccccc||ccc}
\textbf{Segment}             & \textbf{HPC System}  & \makecell{ \textbf{Nodes}} & \makecell{\textbf{Sensors}} & \makecell{\textbf{Data} \\ \textbf{Points}}   & \textbf{Length} & \makecell{\textbf{Sampling} \\ \textbf{Interval}} & \makecell{\textbf{Feature} \\ \textbf{Sets}} & \makecell{$w^l$} & \makecell{$w^s$} \\ \hline
\textbf{Fault}     & \ethz Testbed & 1 & 128 & $\sim$1000k & 16d    & 1s & $\sim$130k & 1m & 10s  \\ \hline
\textbf{Application} & \sng  & 16 & 52 & $\sim$1000k  & 1d     & 1s & $\sim$250k & 30s & 5s  \\ \hline
\textbf{Power}   & \cmuc  & 1 & 47 & $\sim$300k  & 8h      & 100ms & $\sim$60k & 1s & 500ms  \\ \hline
\textbf{Infrastructure} & \cmuc &  148 & 31 & $\sim$450k        & 16d    & 10s & $\sim$75k & 5m & 1m \\ \hline
\textbf{Cross-Arch} & Multiple &  3 & (52,46,39) & $\sim$300k        & 1d    & 1s & $\sim$140k & 30s & 2s 
\end{tabular}
\end{center}
\end{table*}

\subsection{Contributions}
\label{section:contributions}
In this work we propose the \emph{Correlation-wise Smoothing} (CS) algorithm, a method to extract knowledge from and compress multi-dimensional time-series data, tailored for data center monitoring and ODA processes. Our particular use case stems from the area of HPC, but our technique is equally applicable to data center installations in general. 

The CS algorithm is simple, lightweight and designed around online operation: it produces low-dimensionality image-like signatures that yield very good performance in many ODA use cases and at the same time are interpretable to the human eye. Moreover, as the signatures have clear scaling properties and semantics, they can be manipulated at will, thus enabling comparability and portability of models across systems, as well as robustness against dataset structure changes. We apply the CS method to HPC-ODA, a collection of diverse HPC monitoring datasets which cover most typical machine learning-based ODA use cases found in large-scale data center facilities, demonstrating its properties and performance: CS performs as well as other state-of-the-art methods, while producing signatures that are up to one order of magnitude smaller and in up to one order of magnitude less time. The HPC-ODA collection is made available with this work for use by the community.

\subsection{Organization}
\label{section:organization}
The paper is organized as follows. In Section~\ref{section:dataset} we introduce the HPC-ODA dataset collection, which we use for our experiments. In Section~\ref{section:algorithm} we then describe the CS method, and in Section~\ref{section:results} we discuss the experimental results obtained with it. In Section~\ref{section:conclusions} we conclude the paper.

\section{The HPC-ODA Dataset Collection}
\label{section:dataset}

Publicly available datasets are rare in the area of ODA, as data center sensor data and logs are typically considered sensitive both due to privacy and operational concerns. To enable this work we therefore introduce HPC-ODA, a comprehensive collection of datasets captured on several HPC installations, and make it openly available to the research community to aid ODA research. HPC-ODA is representative of HPC centers and their typical workloads, and contains a large number of sensors covering the operations of several systems. In the following, we explain its purpose and composition.

\subsection{Overview of the Dataset Collection}
HPC-ODA is composed of five \emph{segments}, which are self-contained datasets, each acquired separately on a particular HPC system and containing monitoring sensor data associated with different components (e.g., CPU cores, compute nodes or cooling systems) at a well-defined time granularity. The structure of the datasets is simple: each sensor's time-series data is stored in a separate CSV file, with each entry being a time-stamp/value pair. Then, each of the segments has an associated ODA use case which is representative of the state-of-the-art of data center operations; these will be discussed in detail in Section~\ref{section:results}. The monitoring frameworks used to acquire the data in HPC-ODA are the \emph{Data Center Data Base} (DCDB)~\cite{netti2019dcdb} and the \emph{Lightweight Distributed Metric Service} (LDMS)~\cite{agelastos2014lightweight}. However, the data captured in HPC-ODA is generic and not framework-dependent. 

The rationale behind the HPC-ODA segment structure is to provide several vertical slices of the monitoring data typically available in a large-scale data center installation, with their different granularities and time scales. While having a production dataset from an entire HPC system --- from the infrastructure down to the CPU level --- at a fine time scale would be ideal, this is not feasible due to the privacy concerns of the data (e.g., user names or allocation times), as well as the sheer amount of storage space required. To avoid such problems, HPC-ODA provides a compact and sanitized monitoring snapshot that can be used as a reference to evaluate ODA approaches requiring fine-grained data. On the other hand, demonstrating the scalability of models over large-scale systems or providing long-term operational data is not in the scope of HPC-ODA. Moreover, we used HPC applications that are representative of production workloads, such as those from the CORAL-2\footnote{CORAL-2 Suite: \url{https://asc.llnl.gov/coral-2-benchmarks}} suite: more details about application configurations and data acquisition can be found in the documentation of HPC-ODA, which is openly available\footnote{HPC-ODA Dataset Collection: \url{https://zenodo.org/record/3701440}} for use by the research community.

\subsection{Structure of the Dataset Collection}
As mentioned earlier, HPC-ODA is composed of five segments with different characteristics, which can be used independently. The specific structure and parameters of each segment are summarized in Table~\ref{table:hpcoda}. Sensor data consists of monitoring metrics that are commonly available in data center systems: CPU performance counters (e.g., from the \emph{perfevent} Linux interface), as well as memory and OS-related metrics (e.g., from the \emph{proc} file system) can be found in the compute node-level segments, whereas the \emph{Infrastructure} segment includes cooling and power-related data (e.g., a rack's inlet cooling water temperature). The five segments are as follows, together with the ODA use cases we selected.

\vspace{2mm}
\subsubsection{Fault} This segment was collected from a single compute node in a testbed HPC system at \ethz Zurich, while it was subjected to fault injection. It contains compute node-level data, as well as the labels for six single-node applications and eight injected faults: each fault has two possible settings and reproduces various software or hardware issues (e.g., CPU cache contention or memory allocation errors). This segment was extracted from a larger public dataset\footnote{Antarex HPC Fault Dataset: \url{https://zenodo.org/record/2553224}} and is used to perform fault classification, with the goal to optimize management decisions upon anomalous states in HPC components~\cite{netti2019machine}.

\vspace{2mm}
\subsubsection{Application} This segment was collected from 16 compute nodes in the \sng\footnote{SuperMUC-NG: \url{https://doku.lrz.de/display/PUBLIC/SuperMUC-NG}} leadership-class HPC system at \lrz, while running one of six multi-node MPI applications, each under three possible input configurations. Data is at the level of each compute node and is paired with the labels of the running applications. This segment is used to perform application classification, allowing to improve performance for running applications as well as spot rogue codes and purge them from the system~\cite{ates2018taxonomist}.

\vspace{2mm}
\subsubsection{Power} This segment was collected from a single compute node in the \cmuc\footnote{CooLMUC-3: \url{https://doku.lrz.de/display/PUBLIC/CoolMUC-3}} production HPC system at \lrz, while running several single-node OpenMP applications each under two possible input configurations. The segment contains both node-level and CPU core-level data and is used to perform power consumption prediction: this enables system tuning (e.g., via changes in CPU frequency) to optimize performance based on predicted workloads~\cite{ozer2019towards}.

\vspace{2mm}
\subsubsection{Infrastructure} This segment was collected from the infrastructure of the \cmuc HPC system at \lrz and contains data from its power distribution and warm-water cooling systems. Data is at the rack level, with some sensors being at the chassis level. We have no knowledge of the specific applications executed by the users on the system. The segment is used to predict the amount of heat removed by the cooling units: this allows to optimize infrastructure-level knobs (e.g., inlet cooling water temperature) in function of environmental or workload changes~\cite{conficoni2015energy}.

\vspace{2mm}
\subsubsection{Cross-Architecture} This is a variant of the Application segment, containing data associated with six applications each under three input configurations. Here, however, we run single-node OpenMP configurations of the applications on three compute nodes, each with a different hardware architecture. The compute nodes belong to the \sng and \cmuc systems, as well as the BEAST testbed system at \lrz, using respectively Intel Skylake, Knights Landing and AMD Rome CPUs. The three architectures respectively provide 52, 46 and 39 compute node-level sensors, as well as the application labels. The segment is used to perform application classification across different architectures.

\section{The Correlation-wise Smoothing Method}
\label{section:algorithm}
In this section we introduce the Correlation-wise Smoothing (CS) method, alongside a series of formal definitions and a description of the baseline methods we use for comparison.

\subsection{Definitions}
\begin{figure*}[t]
\centering
\includegraphics[width=0.98\textwidth,trim={0 0 0 0}, clip=true]{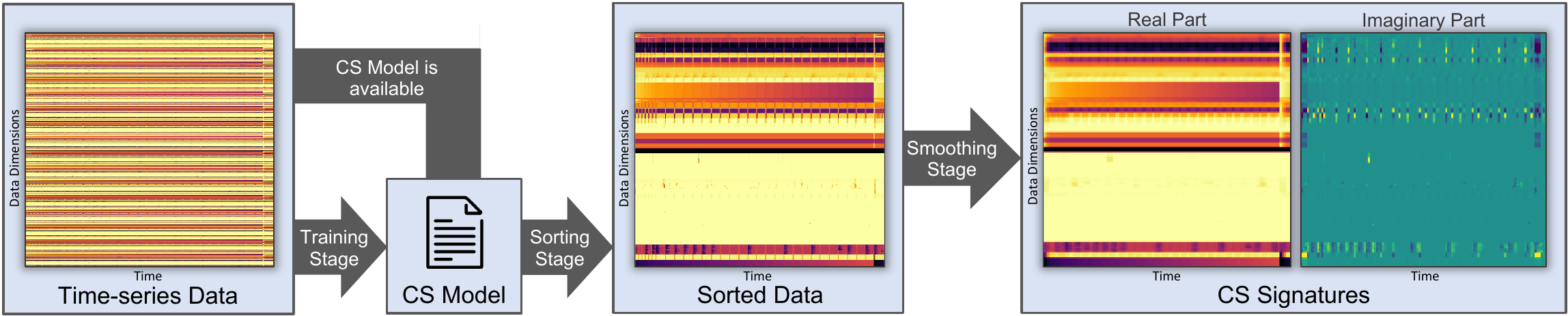}
\caption{The three stages of the CS algorithm. On the left we show a heatmap of the raw sensor data from 16 compute nodes during a run of AMG, from the HPC-ODA Application segment. In the center, the same data is depicted after applying the sorting stage of the CS algorithm. On the right side we show the resulting signature heatmaps using 160 blocks. Darker colors correspond to higher values and each column corresponds to a separate signature.}
\label{method:csflowchart}
\end{figure*}

The problem at hand consists of computing compact representations (\emph{signatures}) from multi-dimensional time-series data originating from a sensor monitoring system, which can be used as feature sets for machine learning and data mining models or simply be visualized. This data can be represented as a \emph{sensor matrix}, in which each row corresponds to a sensor (e.g., power or temperature) in the system and each column to a time-stamp --- each matrix element is thus a numerical reading for a sensor at a specific time-stamp. The sensor matrix $S$ has $n$ rows, as many as sensors, and $t$ columns, as many as individual time-stamps, with indexes starting from 1, and can be visualized as a \emph{heatmap}, as shown in Figure~\ref{method:csflowchart}. For the sake of simplicity, we assume that the sensors in $S$ are time-aligned and have the same sampling rate: this is not necessarily true for real datasets, and an interpolation pre-processing step may be required to align the data.

Seen formally, a \emph{signature method} is a function $Sig()$ that takes as input a sub-matrix $S^w$ of the sensor matrix $S$ and that gives as output a column vector $s^w$. The sub-matrix $S^w$ has $n$ rows and $w^l$ columns: this last parameter is the time \emph{aggregation window} in samples, and in order for the signature method to be useful, the relation $l << n \cdot w^l$ must be satisfied, where $l$ is the final length of the signature. Depending on the method, $l$ can either be an arbitrary parameter or a function of $n$ and $w_l$. We also define $w^s$, which is the \emph{step} between successive aggregation windows. In other words, the signature method $Sig()$ performs aggregation of the sensor data in $S^w$, returning a compressed representation that is more compact than the data it was computed from. In the data center domain, $S^w$ will contain sensor data associated with a system component, such as a compute node, a CPU core or a GPU accelerator; however, $S^w$ can also contain data associated with an entire distributed system or a certain logical partition, such as an HPC user job. The resulting signature $s^w$ can be seen as a vector of coefficients describing its status.

\subsection{Baseline Methods}
\label{section:baseline}

We have identified three baseline signature methods for system monitoring data in the literature. These have been applied to many ODA use cases and are suitable for use in production data centers. While we could select more sophisticated data mining algorithms as baselines, this would be unrealistic as they would never be usable, due to factors such as overhead and latency. The methods are as follows:

\begin{itemize}
    \item \textbf{Tuncer}~\cite{tuncer2018online}: for each sensor row in $S^w$, a series of statistical indicators is computed from its $w^l$ samples. These are the \emph{mean}, \emph{standard deviation}, \emph{minimum}, \emph{maximum}, 5th, 25th, 50th, 75th and 95th \emph{percentiles}, \emph{sum of changes} and \emph{absolute sum of changes}. The last two metrics were used in place of the \emph{skewness} and \emph{kurtosis}, as they led to better performance. The signature's size is $l = n \cdot 11$.
    \item \textbf{Bodik}~\cite{bodik2010fingerprinting}: similarly to the Tuncer method, a series of statistical indicators is computed for each row in $S^w$. In this case only the \emph{minimum}, \emph{maximum} and 5th, 25th, 35th, 50th, 65th, 75th and 95th \emph{percentiles} are used. The signature's size is $l = n \cdot 9$.
    \item \textbf{Lan}~\cite{lan2009toward}: each sensor row in $S^w$ is sub-sampled to a fixed size $w^r$ (smaller than $w^l$) using a mean filter and is then concatenated to the final signature. The original method simply \emph{flattened} the sub-matrix $S^w$ to the vector $s^w$ and employed an additional PCA step for outlier detection. We omit the latter and introduce a sub-sampling step for scalability reasons. The signature's size is $l = n \cdot w^r$.
\end{itemize}

\subsection{Correlation-wise Smoothing}
The methods above do not exploit the relationships between sensors to improve the signatures' quality, which is especially relevant in the case of system monitoring data: in particular, HPC systems are traditionally used to execute large-scale parallel, homogeneous codes, leading to strong correlations between the sensors in the associated compute nodes. Even without this assumption, sensors in system components tend to be highly correlated due to their performance relationships, and sometimes the vast amount of available metrics simply supplies redundant information that can be pruned~\cite{gimenez2017scrubjay}. Under these assumptions, we propose the Correlation-wise Smoothing (CS) method: it is able to perform aggregation \emph{across} sensors, in turn allowing users to enforce arbitrary signature sizes. The CS method's algorithm is designed for lightweight online operation with up to thousands of dimensions and consists of three stages, which we describe in the following. Figure~\ref{method:csflowchart} shows an overview of the algorithm's flow when applied to data from the HPC-ODA Application segment.

\vspace{2mm}
\subsubsection{Training Stage}
In this stage the CS algorithm learns the relationships between data dimensions, using a sensor matrix $S$ that contains historical data characterizing the behavior of the underlying system component --- in our example in Figure~\ref{method:csflowchart}, this includes 16 compute nodes running the AMG application for a total of {\raise.17ex\hbox{$\scriptstyle\sim$}}800 data dimensions. As it can be seen, $S$ is noisy and provides little visual information. Our intuition is that the rows in $S$ can be \emph{permutated}, such that sensors which are correlated to one another will be grouped together, both improving visualizability and suitability for compression. To this end we use the following metrics:

\begin{equation}
\begin{aligned}
    \rho_{S_i,S_j} = \frac{cov(S_i, S_j)}{\sigma_{S_i}\sigma_{S_j}} + 1 \qquad \rho_{S_i} = \frac{1}{n-1}\sum_{j \neq i} \rho_{S_i,S_j}
\end{aligned}
\label{eq:sorting}
\end{equation}

In Equation~\ref{eq:sorting}, $\rho_{S_i,S_j}$ is the Pearson correlation coefficient between rows $i$ and $j$ of $S$, which is computed as the ratio between their covariance $cov(S_i, S_j)$ and the product of the respective standard deviations $\sigma_{S_i}$ and $\sigma_{S_j}$. Coefficients are shifted by $1$ to ensure that they are non-negative and in the interval $[0, 2]$. Then, the \emph{global} correlation coefficient $\rho_{S_i}$ is defined as the sum of all correlation coefficients with respect to row $i$: this metric quantifies the relevance of a certain dimension in the entire dataset, and how well it can describe the system's status. Using these two metrics we compute a row permutation vector $p$ as described in Algorithm~\ref{fig:alg1}: starting from the row with the maximal $\rho_{S_i}$, rows are selected iteratively as those maximizing the product between $\rho_{S_i}$ and the correlation coefficient with the latest row added to $p$. 

This type of ordering is generic and interpretable: the sensors at the beginning of $p$ are those that meaningfully describe the system's status and that have an overall positive correlation with other sensors. Sensors at the middle of $p$ have little correlation with other sensors and are akin to noise. Finally, sensors at the end of $p$ are again descriptive of the system's status, but are negatively correlated with those at the beginning. On top of the permutation vector $p$, the training stage also computes the lower and upper bounds of each row to enable min-max normalization of the data: these data structures compose a \emph{CS model}, which can be stored and re-used for the subsequent stages of the algorithm. The computational complexity of this stage is dominated by the correlation matrix computation and amounts to $O(n^2t)$.

\begin{figure}[b]
    \centering
 \begin{algorithm}[H]
 \caption{Correlation-wise Smoothing - Training Stage.}
 \label{fig:alg1}
 \begin{algorithmic}[1]
    \fontsize{8.5}{8.5}\selectfont
    \renewcommand{\arraystretch}{1.2}
    \renewcommand{\algorithmicrequire}{\textbf{Input:}}
    \renewcommand{\algorithmicensure}{\textbf{Output:}}
    \REQUIRE the coefficients $\rho_{S_i,S_j}$ and $\rho_{S_i}$ for $0 < i,j \leq n$
    \ENSURE  a permutation vector $p$
    \STATE \textbf{vector of int} $p$ $ \gets \varnothing $
    \STATE \textbf{set of int} $d$ $ \gets (1, 2, ..., n) $
    \STATE next $\gets argmax(\rho_{S_k}$ for $k \in d$)
    \STATE $d$.remove(next)
    \STATE $p$.append(next)
    \WHILE {$d$ is not empty}
    \STATE next $\gets argmax(\rho_{S_k, S_{next}}\cdot\rho_{S_k}$ for $k \in d$)
    \STATE $d$.remove(next)
    \STATE $p$.append(next)
    \ENDWHILE
    \RETURN $p$
    \end{algorithmic}
    \end{algorithm}
\end{figure}

\vspace{2mm}
\subsubsection{Sorting Stage}
The second and third stages are performed any time a signature must be computed from a sub-matrix $S^w$ of $S$. Here, min-max normalization is applied to the sensor data and the permutation vector $p$ is used to sort the rows of $S^w$. As shown at the center of Figure~\ref{method:csflowchart}, simply re-arranging the rows in $S$ brings clear visual patterns to the surface, vastly improving the quality of visualization. This is due to the fact that highly-correlated sensors follow the same trends over time, which are highlighted by grouping them together. The processed heatmaps can be treated as images, opening up the possibility to leverage a whole branch of image processing techniques for data center monitoring. By default the training stage of the CS algorithm is performed only once, re-using the CS model for all new sensor data, thus greatly reducing its footprint. However, the algorithm can be configured to repeat the training stage whenever required: this can be useful when dealing with data from multiple components of a large-scale system (e.g., compute nodes), whose relationships and correlations change over time depending on how they are used. The computational complexity of this stage is dominated by the min-max normalization, amounting to $O(w^ln)$.

\vspace{2mm}
\subsubsection{Smoothing Stage}
The last stage performs a \emph{smoothing} operation over the sorted $S^w$ sub-matrix and creates the final signature, which is composed of an arbitrary number $l$ of complex-valued elements (or \emph{blocks}), each aggregating a partially overlapping range of sensors:

\begin{equation}
    b_i = 1 + \lfloor (i-1) \cdot n/l \rfloor  \qquad e_i = \lceil i \cdot n/l \rceil
\label{eq:blockranges}
\end{equation}

In Equation~\ref{eq:blockranges}, $b_i$ and $e_i$ express the indexes of the first and last sensors aggregated in each block $i$ in the range $[1,l]$. This blocking scheme has two properties: first, each block refers to a specific set of sensors, with their ordering being an importance indicator. Second, when the modulo \(n\%l\) is not zero, a corresponding amount of blocks will be extended by 1 sensor; these larger blocks are distributed uniformly over the signature due to the periodicity of the modulo operation. Finally, as the set of raw sensors belonging to a block is clearly defined, root cause analysis is simplified. The values of each block are then computed as follows:

\begin{equation}
\begin{aligned}
    Re(s^w_i) & = \frac{1}{w^l(e_i-b_i+1)}\sum_{j=b_i}^{e_i} \sum_{k=1}^{w^l} S^w_{j,k} \\
    Im(s^w_i) & = \frac{1}{w^l(e_i-b_i+1)}\sum_{j=b_i}^{e_i} \sum_{k=1}^{w^l} S^{w'}_{j,k}
\end{aligned}
\label{eq:smoothing}
\end{equation}

In Equation~\ref{eq:smoothing}, $S^{w'}$ is the matrix of row-wise first-order derivatives computed from $S$ using backward finite differences and corresponding to the time window of $S^w$. The real part of each block $s^w_i$ thus contains the average value of the sensors in block $i$, in the time range described by $S^w$; the imaginary part, instead, contains the corresponding average first-order derivatives. The right side of Figure~\ref{method:csflowchart} shows two heatmaps containing the real and imaginary parts of the signatures computed from the raw AMG sensor data: each column represents a separate signature whose real part describes the \emph{static} properties of the system, while the imaginary part describes its \emph{dynamic} properties. In particular, the heatmap of the real signature parts presents itself as a smoothed version of the sorted $S$ matrix, both on the time and sensor axes. The signatures can be scaled at will using traditional image processing algorithms, ensuring comparability as well as portability, while still allowing users to choose the resolution (defined by $l$, $w^l$ and $w^s$) that they deem appropriate. Further, more aggressive compression is possible: as the central signature coefficients represent the least insightful sensors in the system, they can be potentially eliminated with minimal loss of information. Due to its normalization step, the CS algorithm cannot be used with monotonic time-series (e.g., energy), but this was not found to be an issue for monitoring data, since time-series of this type can be transformed using backward finite differences. The complexity of the smoothing stage amounts to $O(w^ln)$. 


\section{Experimental Evaluation}
\label{section:results}
In this section we present our experimental evaluation of the CS method carried out using HPC-ODA, as well as its performance against the three baseline methods. Each of the experiments targets some of the requirements stated in Section~\ref{section:introduction}, in order to demonstrate the suitability of the CS method for production ODA use cases.

\subsection{Experiment Setup}

\vspace{2mm}
\subsubsection{Methodology}
We state all of the ODA use cases introduced in Section~\ref{section:dataset} as classification and regression machine learning tasks as follows:

\begin{itemize}
    \item \textbf{Fault}: we use signatures to detect the eight possible injected faults, as well as healthy operation (classification).
    \item \textbf{Application}: we use signatures to distinguish the six applications, or idle operation, in each node (classification).
    \item \textbf{Power}: we use signatures to predict the average compute node power consumption (at the outlet) in the next 3 samples, or $\sim$300ms (regression).
    \item \textbf{Infrastructure}: we use signatures to predict the average amount of heat removed from each rack over the next 30 samples, or $\sim$5m (regression).
    \item \textbf{Cross-Architecture}: we use signatures to distinguish the six applications on each architecture (classification).
\end{itemize}

For each selected signature method, we process each segment in HPC-ODA using specific $w^l$ and $w^s$ values as described in Table~\ref{table:hpcoda}, with the size of each resulting dataset under the \emph{feature sets} column. The order of the feature sets (i.e., the signatures) in each dataset is then shuffled, and we apply 5-fold cross-validation to them using a stratified K-fold strategy: according to this scheme, 4 of the 5 uniformly-sized folds are used for training and 1 for testing, evaluating all possible combinations. The input to a model is always a single signature vector, and never a signature heatmap. As machine learning model we employ a random forest (with 50 estimators and using the Gini impurity to evaluate the quality of splits), due to its effectiveness in many ODA use cases as well as its robustness against over-fitting~\cite{tuncer2018online, ates2018taxonomist}. In a subset of our tests, we also employ a multi-layer perceptron (with 2 hidden layers each having 100 neurons and using the rectified linear unit as activation function). The models are trained as classifiers or regressors depending on the problem at hand, and we use two metrics to evaluate machine learning performance: for classification problems we use the \emph{F1-score}, which is computed as the harmonic mean between the \emph{precision} and \emph{recall} metrics. For regression problems we instead use the \emph{Normalized Root Mean Square Error} (NRMSE): in order to show both metrics in a comparable \emph{higher-is-better} fashion, referred to as \emph{ML score}, we use the formula $\mathit{NRMSE}_c = 1 - \mathit{NRMSE}$ to complement the NRMSE. Experiments are repeated 5 times and we show average results.

\vspace{2mm}
\subsubsection{Similarity Metrics}
In some experiments we also analyze the fidelity of the compression generated by the CS method, and how it changes with different signature sizes. Several metrics are available in the literature, such as the \emph{Kullback-Leibler} (KL)~\cite{kullback1951information} or the \emph{Jensen-Shannon} (JS) \emph{divergence}~\cite{dagan1997similarity}: however, these are heavily affected by the curse of dimensionality, since they rely on the joint probability distributions of the datasets' dimensions. A generalized form of the JS divergence exists~\cite{aslam2007query} and can be applied to an arbitrary number of dimensions, but it operates by comparing them against one another and not in separate sets. To circumvent this issue we leverage the property that data treated with the CS method's sorting stage can be visualized in image-like form, as shown in Figure~\ref{method:csflowchart}: as such, dimensions in the original data can be directly mapped to those of the CS signatures and there is no need to use $n$-dimensional probability distributions to compute the divergence metrics. Instead, we collapse the probability distributions to a 2-dimensional space, where the vertical axis captures the different dimensions and the horizontal axis their values, and use the following JS divergence equation:

\begin{equation}
\begin{aligned}
    JS(P_d || P_s) & = H\bigg(\frac{P_d + P_s}{2}\bigg) - \frac{H(P_d) + H(P_s)}{2}
\end{aligned}
\label{eq:jsdiv}
\end{equation}

In Equation~\ref{eq:jsdiv}, $P_d$ and $P_s$ represent the 2-D probability distributions of the original uncompressed data (treated with the CS method's sorting stage) and that of the CS signatures respectively, while $H()$ is the Shannon entropy function. In practical terms, $P_d(v, y)$ or $P_s(v, y)$ signifies the \emph{probability of value $v$ on dimension $y$}, which is computed from the marginal probability distribution of dimension $y$ and divided by $n$, to ensure that $P_d$ and $P_s$ are in turn probability density functions. Furthermore, we apply nearest-neighbor interpolation to the CS data along its vertical axis to ensure that it has the same number of dimensions as the original. This process is performed twice, once for the real components of the CS signatures and once for the imaginary ones, against the uncompressed first-order derivatives of the original data, and we then compute an average JS divergence value lying in the $[0, 1]$ range. For this metric, lower values indicate higher similarity between the datasets being compared. 

\vspace{2mm}
\subsubsection{Environment}
All experiments were carried out in a Python 3.6 environment, using the scikit-learn 0.20.3 and numpy 1.16.2 packages. The baseline methods described in Section~\ref{section:baseline}, as well as the CS method, were implemented in Python leveraging native numpy functions as much as possible to optimize performance. All of our implementations, together with the scripts needed to reproduce all of the results obtained against HPC-ODA in this section, are included in the dataset distribution as a self-contained Python framework. Finally, the machine on which the tests were conducted is equipped with two 14-core Intel Haswell Xeon E5-2697 v3 CPUs, 64GB of RAM and runs Linux SLES15 SP1.

\subsection{Machine Learning Performance}
\label{section:mlperformance}

\begin{figure}[t]
\centering
\captionsetup[subfigure]{}
    \subfloat[Dataset generation (bottom bar) and cross-validation times (top bar).]{
    \includegraphics[width=0.475\textwidth,trim={0 10 0 10}, clip=true]{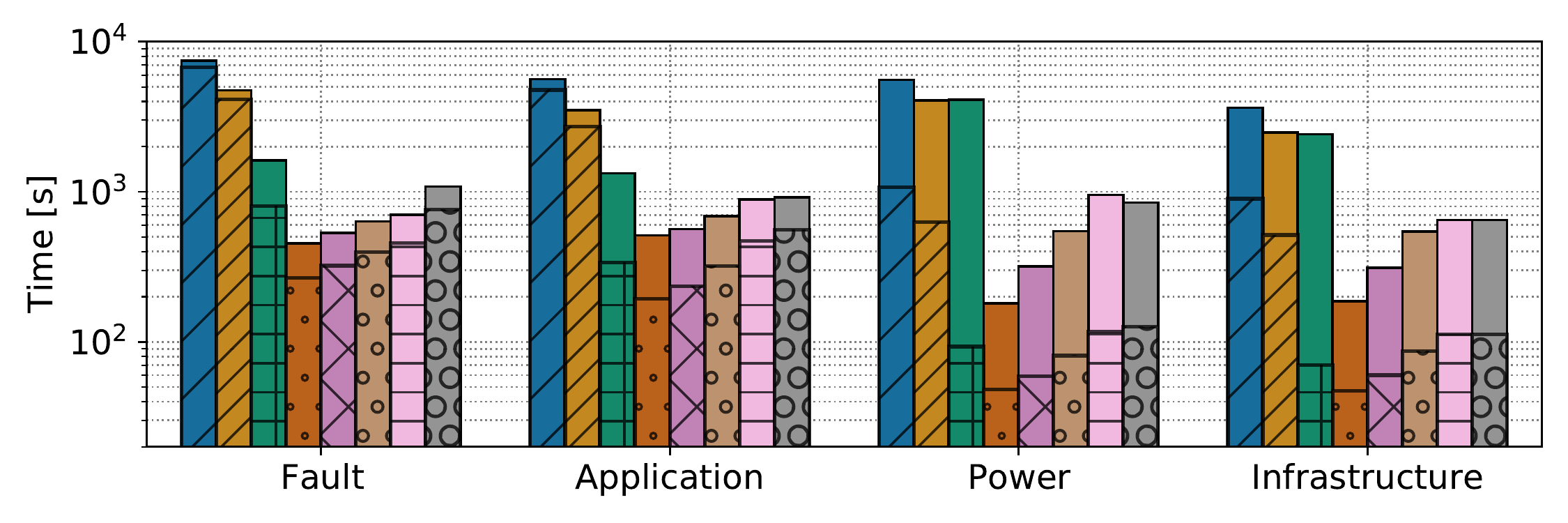}
  	}
  \\	
      \subfloat[Sizes of the signatures.]{
    \includegraphics[width=0.475\textwidth,trim={0 10 0 10}, clip=true]{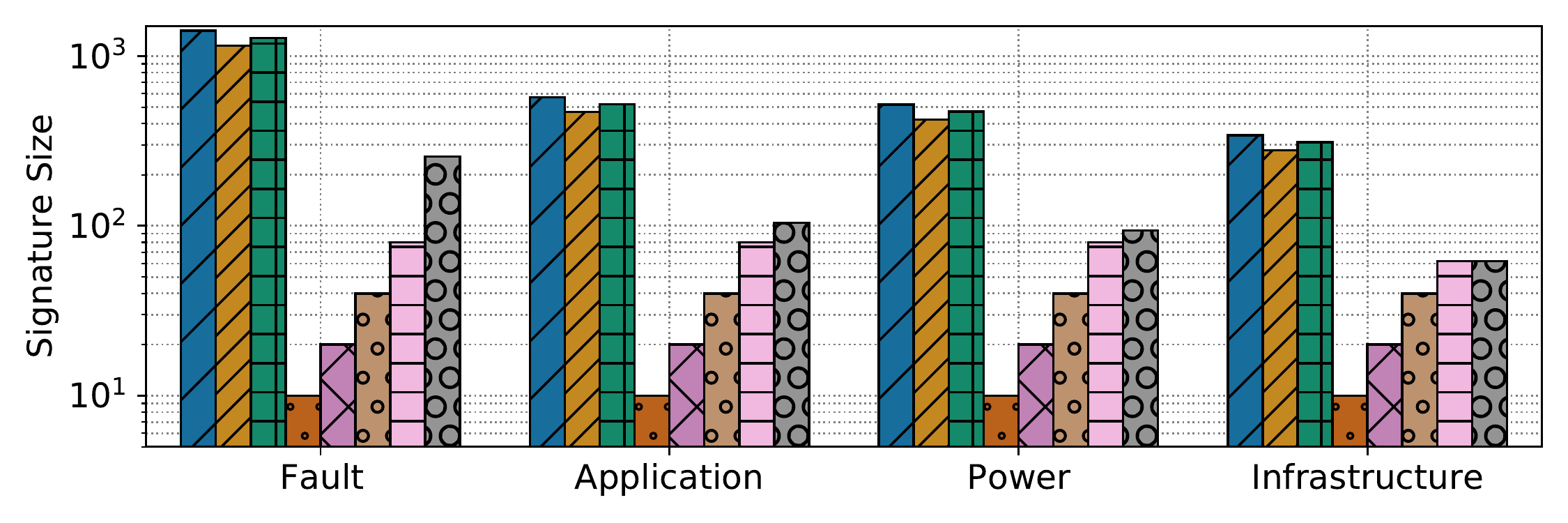}
  	}
  \\
  \subfloat[Machine learning scores.]{
    \includegraphics[width=0.475\textwidth,trim={0 10 0 10}, clip=true]{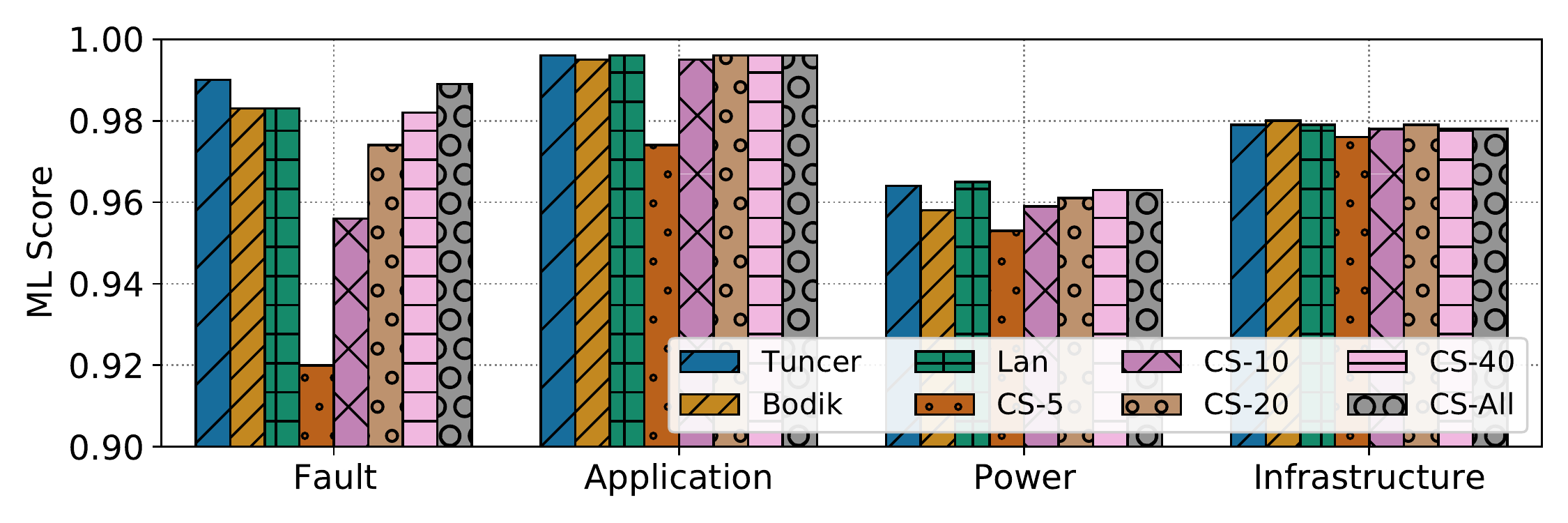}
  }
  \caption{Average testing times (a), resulting signature sizes (b) and machine learning scores (c) in terms of \emph{F1-score} (Fault and Application) and \emph{1-NRMSE} (Power and Infrastructure) for each method.}
  \label{results:mlperf}
\end{figure}

Here we show the results of 5-fold cross-validation applied to the first 4 segments of the HPC-ODA collection using the baseline methods in Section~\ref{section:baseline} as well as the CS method with 5, 10, 20, 40 or \emph{all} (as many as the sensors in the segment) blocks; in this experiment we use random forests as models. Results are shown in Figure~\ref{results:mlperf}: in Figure~\ref{results:mlperf}a we show the testing time for each signature method on each segment, separated between the dataset generation and cross-validation phases, while in Figure~\ref{results:mlperf}b we show the corresponding signature sizes and in Figure~\ref{results:mlperf}c the ML scores. Among the baseline methods, the Tuncer one tends to achieve the best accuracy, but it is consistently the worst performer in terms of testing time due to its statistical computations. The Bodik method shows a similar behavior, while the Lan method is a lightweight alternative to the two, with slightly better testing times. 

We see that the CS method achieves the same ML scores as the baseline methods with ease; for some segments this is only achieved using a large amount of blocks (e.g., Fault), whereas for other segments even using only 5 blocks suffices (e.g., Infrastructure). This is dependent on the specific problem at hand and on its reliance on the original sensor data's resolution: fault classification (i.e., Fault) is dependent on the exact values of certain error counters, while removed heat prediction (i.e., Infrastructure) can be accurate even when using only averages of the system's temperature and power consumption. In all cases, however, these results are obtained with signature sizes that are up to one order of magnitude smaller than those associated with the baseline methods. This is reflected in the testing times, where both dataset generation and cross-validation (especially for regression tasks) are up to one order of magnitude faster when using the CS method. The implications of these results are significant: system administrators can evaluate the performance of ODA models using the CS method with varying numbers of blocks, and then freely choose the configuration that yields the best compromise in function of runtime constraints. Moreover, they can also train models using low-resolution signatures and then feed down-scaled high-resolution signatures to them (or do the opposite), allowing to compute a single CS signature per HPC component that can then be scaled and fed into different ODA models according to their needs. The near-optimal ML scores of the CS method are comparable with those of the baseline methods and satisfy the \emph{Performance} requirement.

\subsection{Quality of Compression}
\label{section:scalingproperties}
 
\begin{figure}[b]
\centering
\captionsetup[subfigure]{}
\subfloat[Jensen-Shannon divergence.]{
\includegraphics[width=0.23\textwidth,trim={10 0 10 0}, clip=true]{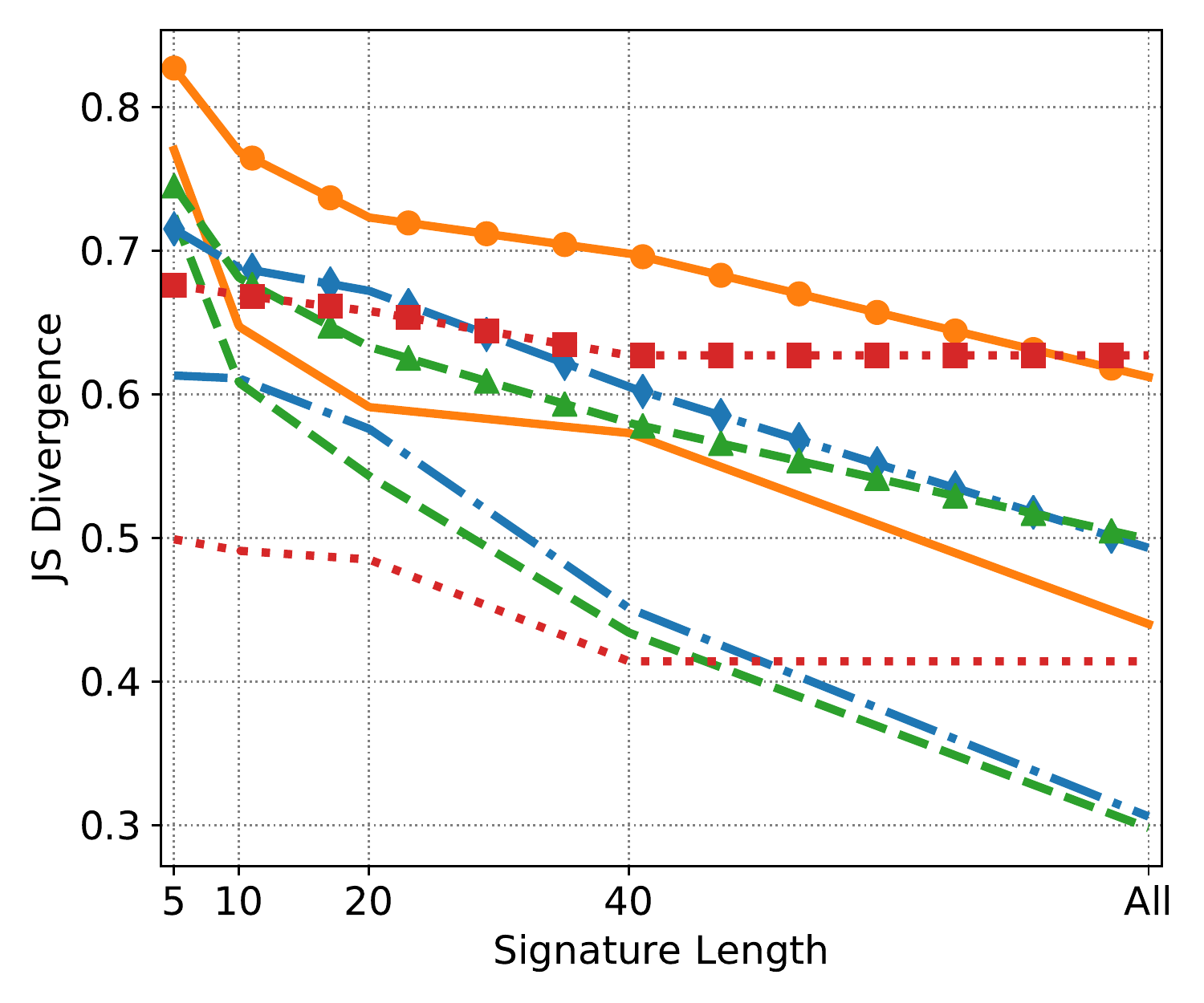}
  }
\subfloat[Machine learning score.]{
\includegraphics[width=0.23\textwidth,trim={10 0 10 0}, clip=true]{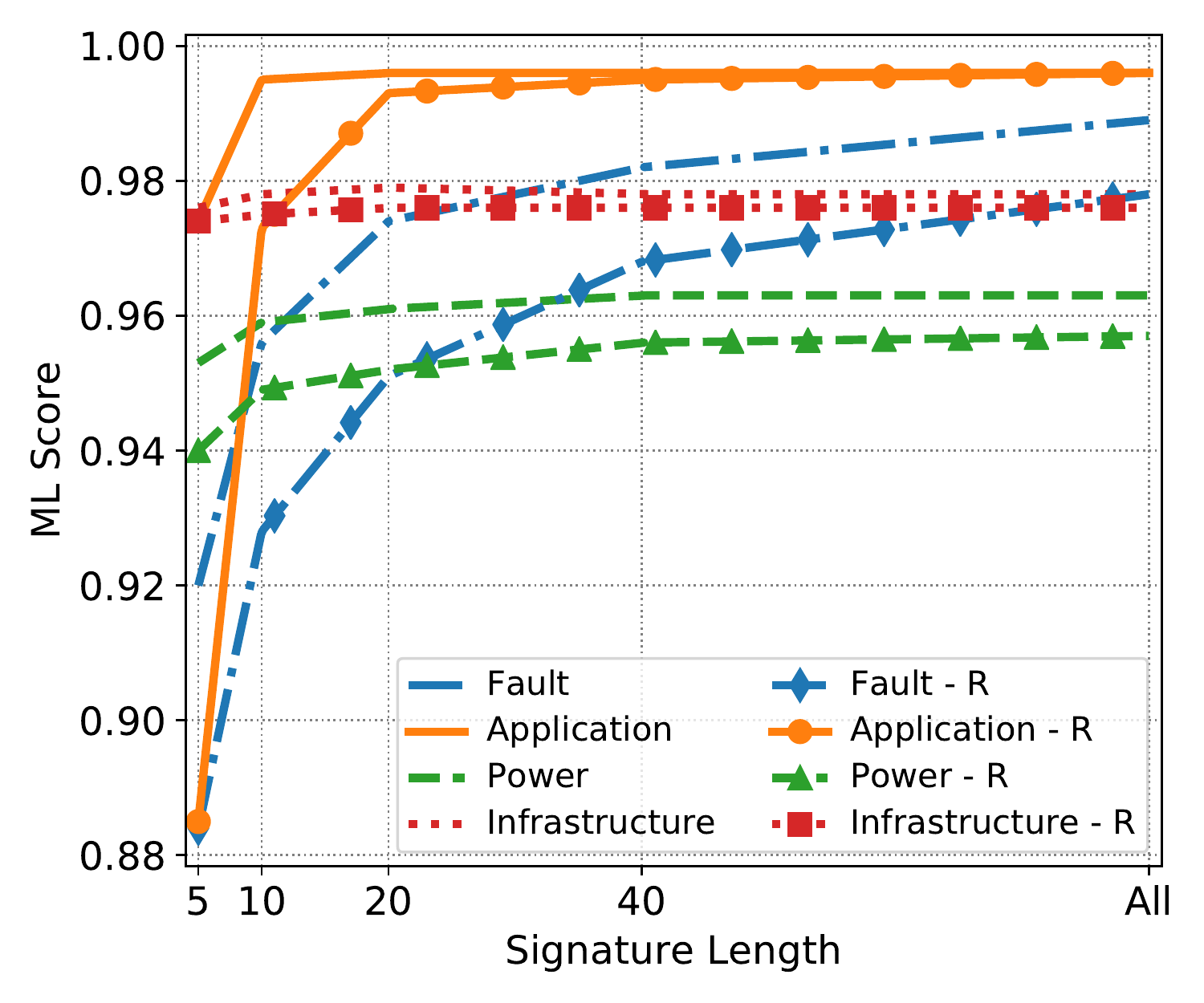}
}
\caption{Jensen-Shannon divergence and machine learning score values for all use cases at different CS signature lengths, as well as when using only the real components (labeled as \emph{-R}).}
\label{results:infloss}
\end{figure}

Here we focus on characterizing the quality of the CS method's compression, again with respect to the first four segments of HPC-ODA. In Figure~\ref{results:infloss} we show the JS divergence of the signature sets processed with the CS method against the original HPC-ODA data, as well as the ML scores in function of $l$. Here, we also show results obtained when using only the real components of the CS signatures, thus omitting the information regarding first-order derivatives. The information in the two plots is complementary: in all cases, an increase in $l$ (or similarly a decrease in $w^l$, which we do not show here for space reasons) corresponds to both higher ML scores and lower JS divergence, suggesting a correlation between the two metrics. This implies that the CS method's compression scales in a stable and predictable way, with more information content being captured at higher block counts. Furthermore, not all HPC-ODA use cases are affected in the same way by changes of $l$: Power and Fault are affected to a greater degree, and their ML scores improve significantly together with $l$. Infrastructure sees only minor effects, with both metrics remaining mostly stable in function of $l$, while in the Application use case the decrease in JS divergence between an $l$ of 5 and 20 translates into a proportional increase in the ML score; after this point, model performance reaches its peak and any further reduction in JS divergence does not have any effect.

Removing the imaginary components of the CS signatures produces interesting results: all use cases see a noticeable impact on JS divergence ({\raise.17ex\hbox{$\scriptstyle\sim$}}0.2 increase), whereas ML scores change according to the features and requirements of the specific problem at hand. The Power and Fault use cases, in particular, see a significant degradation in the ML score, whereas Application is affected only for values of $l$ lower than 20. Infrastructure, on the other hand, exhibits the same ML score in both scenarios. In general, since the imaginary components capture trend-related information and allow to distinguish states that have the same static (i.e., average) information, their relevance depends on the nature of the data as well as its granularity. It should also be noted that in all cases both the JS divergence and ML score remain monotonic in function of $l$, even when removing the imaginary components. With its clear compression properties and its remarkable model performance even at very low values of $l$, the CS method satisfies the \emph{Compression} requirement.

\subsection{Scalability and Computational Complexity}
\label{section:footprint}

\begin{figure}[t]
\centering
\captionsetup[subfigure]{}
\subfloat[Aggregation window.]{
\includegraphics[width=0.23\textwidth,trim={10 0 12 0}, clip=true]{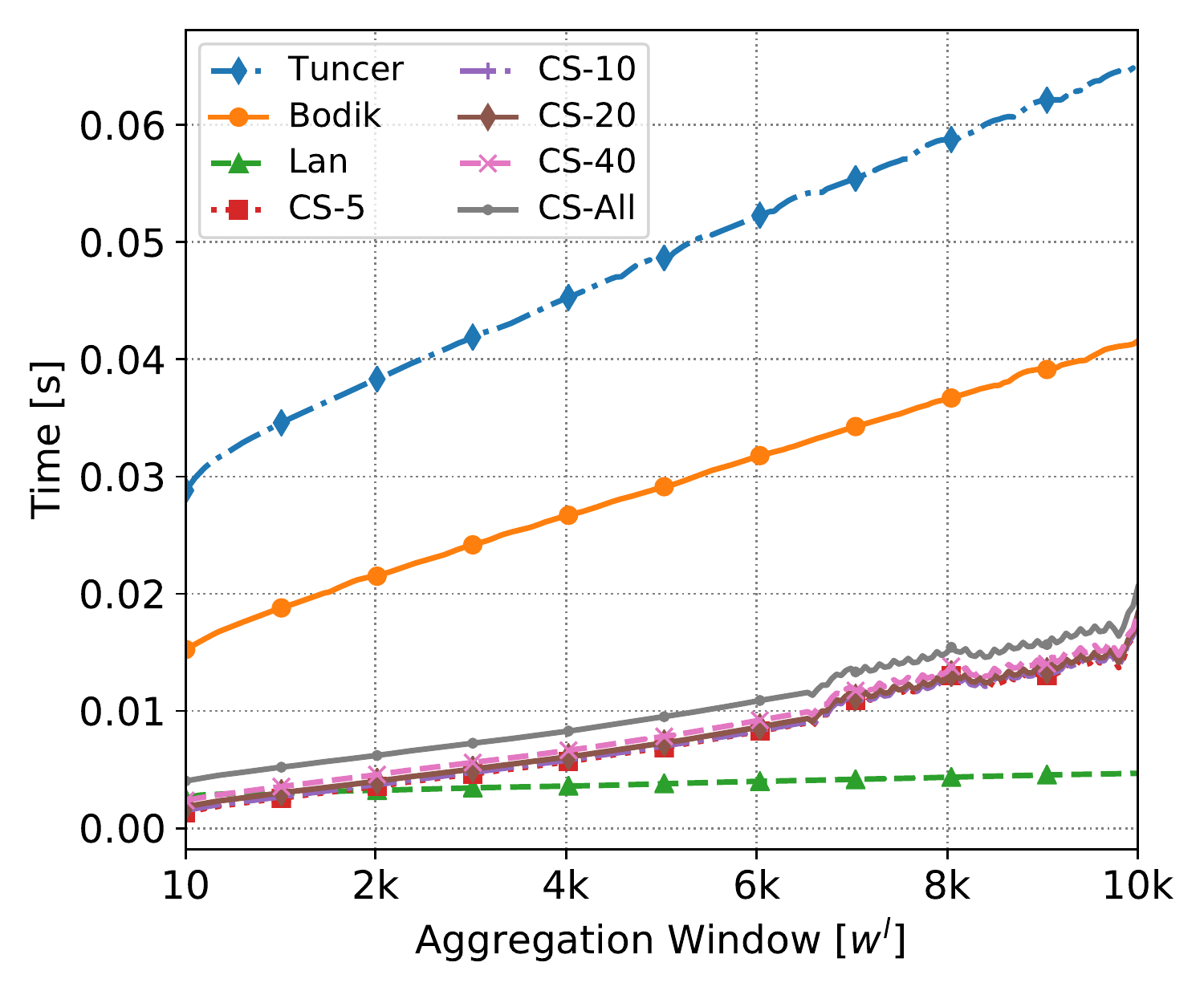}
  }
\subfloat[Number of dimensions.]{
\includegraphics[width=0.23\textwidth,trim={10 0 12 0}, clip=true]{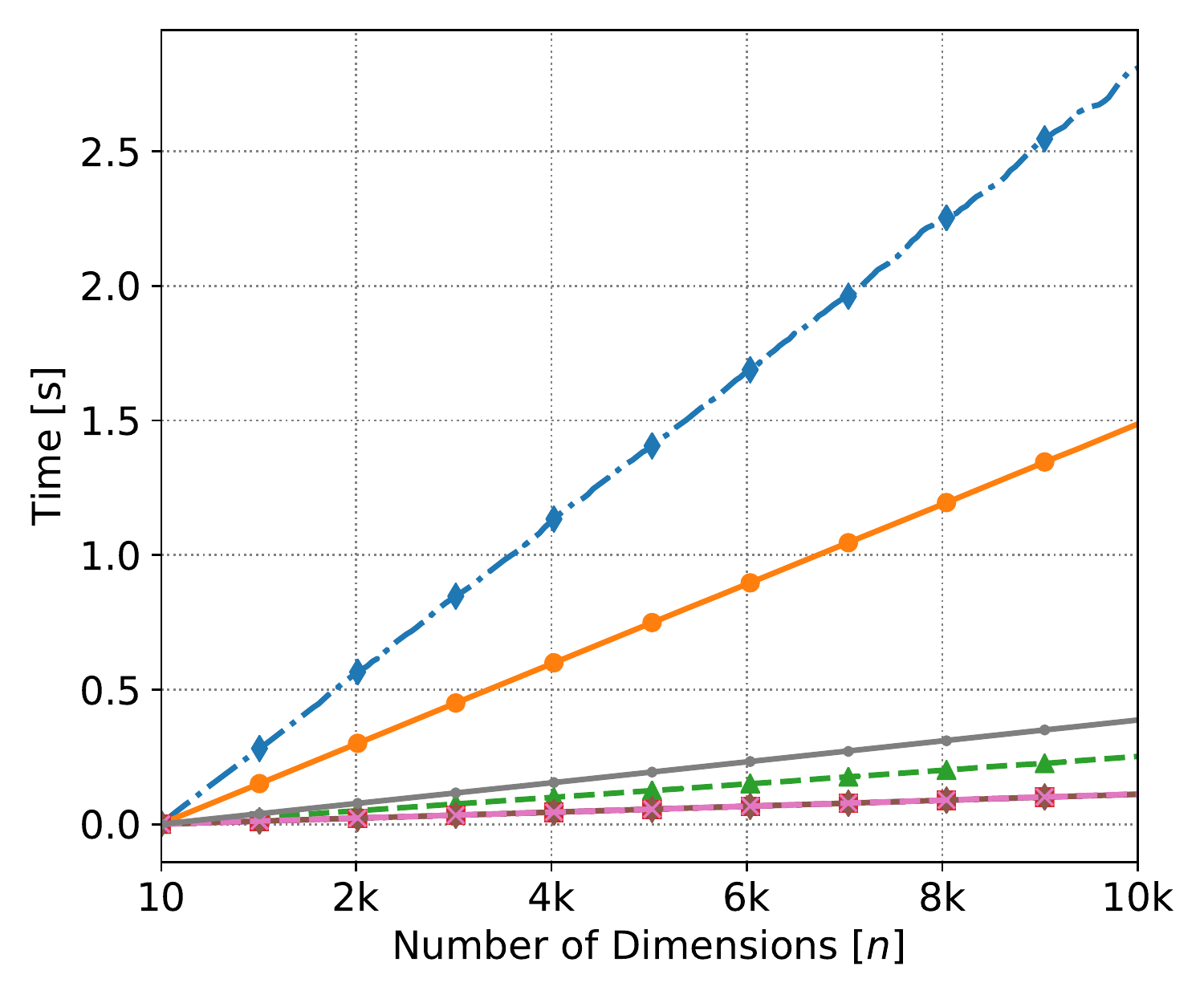}
}
\caption{Time to compute a signature for all methods, in function of the aggregation window \(w^l\) (a) or the number of dimensions \(n\) (b). When varying \(w^l\), \(n\) is fixed to 100 and vice-versa.}
\label{results:footprint}
\end{figure}

We now characterize the CS method's scalability, in terms of the time required to compute a signature, and compare it to its theoretical computational complexity. To this end we generated a series of random \(S^w\) matrices with increasing aggregation window lengths (\(w^l\)) and numbers of dimensions (\(n\)). We then computed signatures for each matrix size and each method, repeating the process 20 times and picking median times. We exclude the CS method's training stage from this evaluation: in most use cases (e.g., in-band ODA in compute nodes) training is performed only once and potentially offline, with a fixed CS model over time. On the other hand, most use cases in which training at every time step is necessary (e.g., system-wide ODA) are out-of-band, and hence overhead and complexity are not a concern. Figure~\ref{results:footprint} shows the observed time to compute a signature for each method in function of \(w^l\) and \(n\). 

It can be seen that the CS method scales linearly both in function of \(w^l\) and \(n\), as expected from its $O(w^ln)$ complexity. The baseline methods exhibit similar behavior, scaling linearly in function of \(n\) as they process each dimension's data independently. On the other hand, the Tuncer and Bodik methods show slightly non-linear behavior in function of \(w^l\), which is due to the computation of percentiles on each dimension, with a complexity of \(O(w^llog(w^l))\). Even though all methods exhibit similar scaling, it can be seen that the CS method performs much better than Tuncer and Bodik at high values of \(w^l\) and \(n\), with Lan being the only exception due to its minimal nature. In particular, the CS method requires one order of magnitude less time to compute a signature from 10k data dimensions compared to Tuncer and Bodik: hence, we deduce that the CS method, like its state-of-the-art counterparts~\cite{tuncer2018online}, could be used in a production environment with negligible impact on performance. Scalability on large-scale systems for the HPC-ODA use cases is ensured by the fact that most of them are in-band (e.g., Fault and Power) and can thus be managed independently on each compute node.

Furthermore, it can be seen that the number of blocks chosen for the CS method has minimal impact on its footprint. Some differences can still be seen, similarly to what was observed in Figure~\ref{results:mlperf}a when processing the HPC-ODA segments: this is a side effect of our Python implementation. There, a higher value of $l$ results in a larger number of calls to native functions and hence worse performance, but we expect it to be negligible in a compiled implementation. Similar considerations can be made for the fluctuations at high values of \(w^l\) in Figure~\ref{results:footprint}a, which are likely due to the interaction between the Python runtime and the system's memory. Thanks to its scalability compared to the baseline methods, the CS method satisfies the \emph{Footprint} and \emph{Scalability} requirements.

\subsection{Fitness for Visualization}
\label{section:visualization}
\begin{figure}[t]
\centering
\captionsetup[subfigure]{}
\subfloat[Kripke.]{
\includegraphics[width=0.23\textwidth,trim={10 0 120 0}, clip=true]{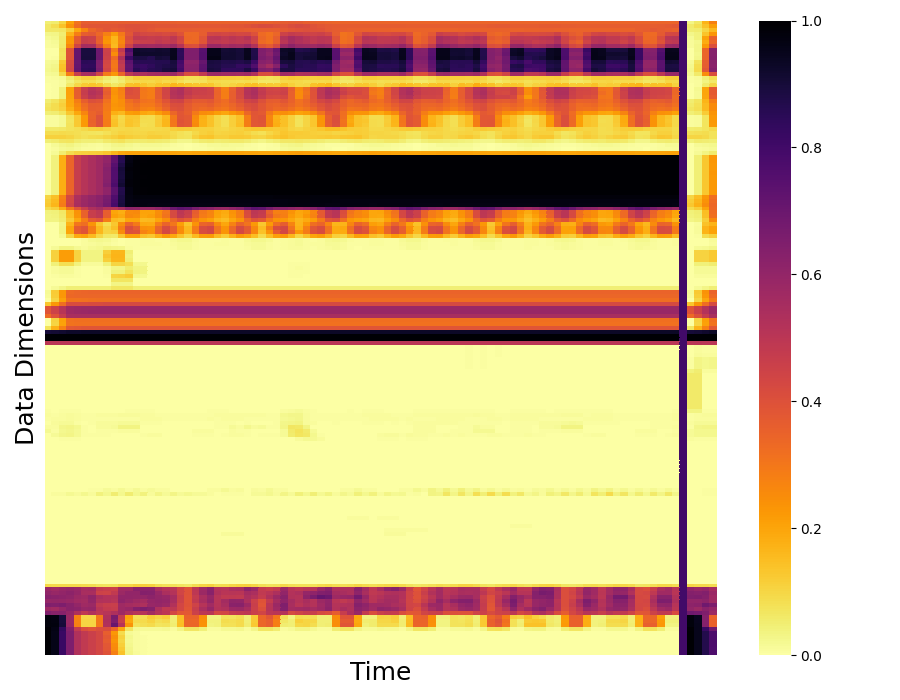}
\includegraphics[width=0.23\textwidth,trim={6 0 120 0}, clip=true]{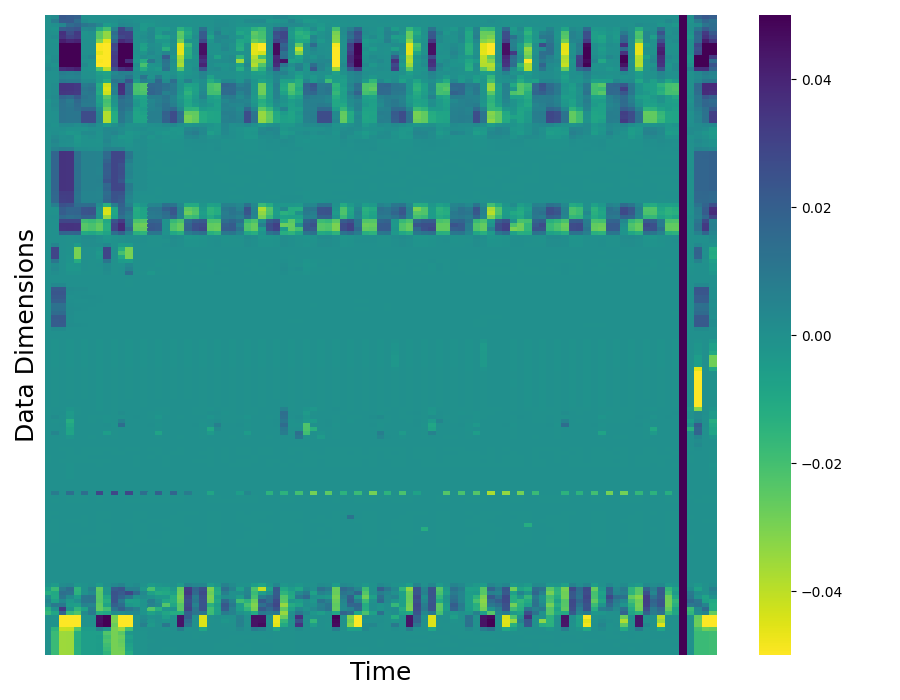}
  }
\\
\subfloat[Linpack.]{
\includegraphics[width=0.23\textwidth,trim={10 0 120 0}, clip=true]{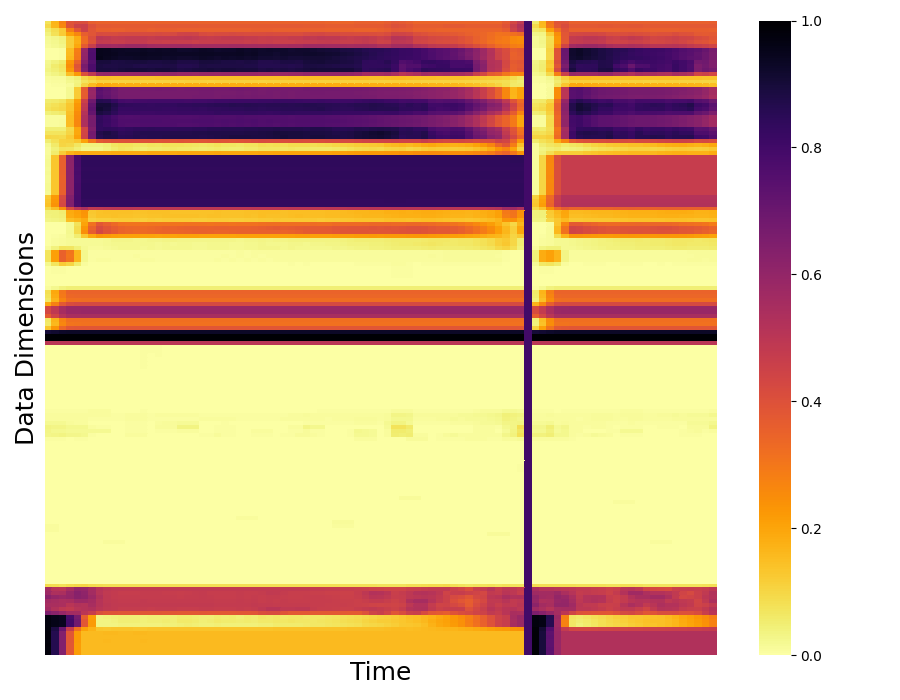}
\includegraphics[width=0.23\textwidth,trim={6 0 120 0}, clip=true]{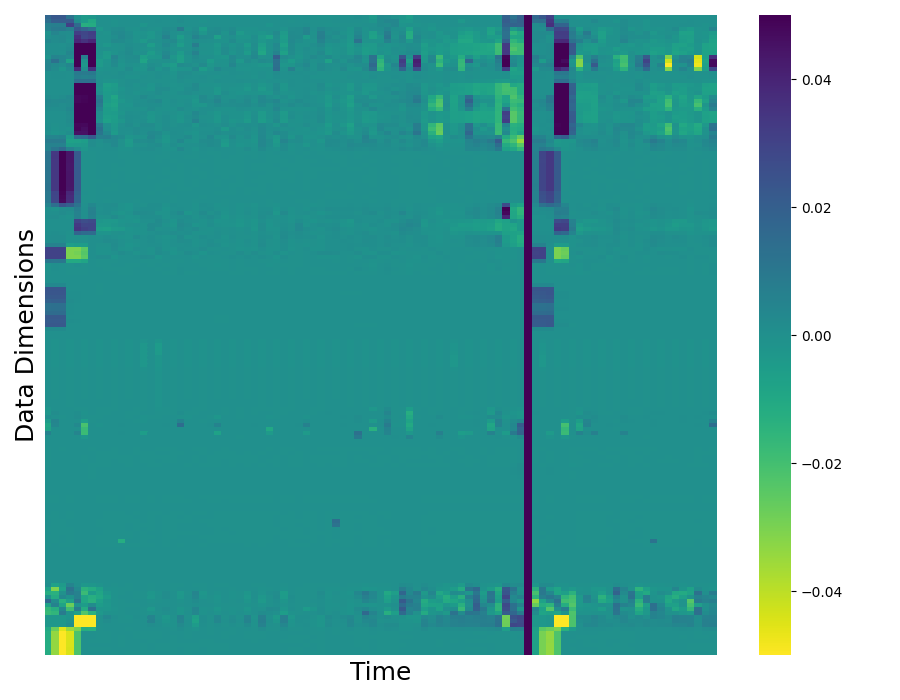}
  }
\\
\subfloat[Quicksilver.]{
\includegraphics[width=0.23\textwidth,trim={10 0 120 0}, clip=true]{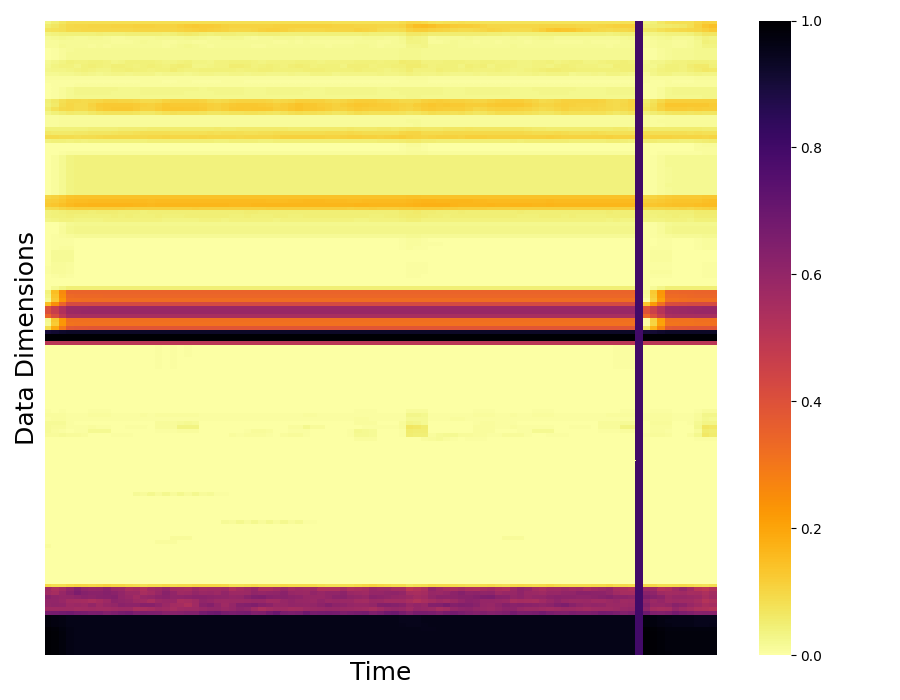}
\includegraphics[width=0.23\textwidth,trim={6 0 120 0}, clip=true]{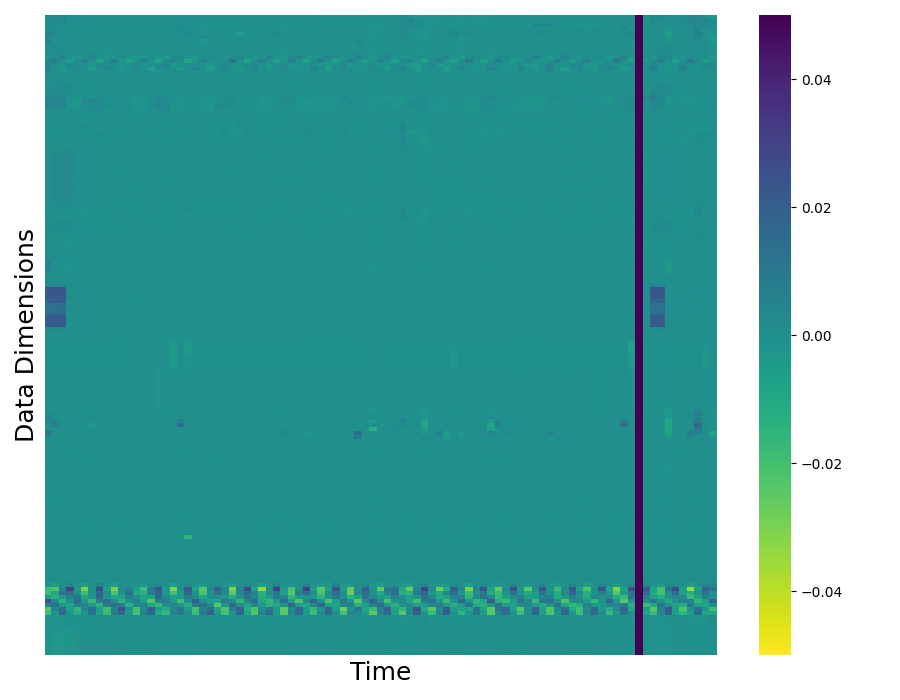}
  }
\caption{Real (left) and imaginary (right) components of the signature heatmaps for three applications using the CS method with 160 blocks. Solid vertical lines indicate the end of a run.}
\label{results:visualizations}
\end{figure}

Having demonstrated the effectiveness of the CS method in various machine learning scenarios, we now demonstrate how it produces signatures that can be easily visualized and interpreted. In Figure~\ref{results:visualizations} we show three signature heatmaps in the same format as those in Figure~\ref{method:csflowchart}: these correspond to Kripke, Linpack\footnote{Linpack Benchmark: \url{https://www.netlib.org/benchmark/hpl/}} and Quicksilver runs from HPC-ODA's Application segment and were computed with data from all 16 compute nodes, resulting in a total of {\raise.17ex\hbox{$\scriptstyle\sim$}}800 sensors. We use 160 blocks, and show both the real and imaginary parts. The applications were executed using one MPI rank per node and as many OpenMP threads as physical cores. Solid vertical lines in the heatmaps separate different runs. 

The three applications can be clearly distinguished and exhibit interpretable performance patterns. For example, Kripke shows a very clear iterative behavior which can be observed in both the real and imaginary components. Linpack, on the other hand, exhibits a constant load on the compute nodes, with a very pronounced initialization phase. Quicksilver, finally, shows very light load on computational resources with small values across all blocks, but a very unique behavior emerges: at the bottom of the imaginary components we can observe a clear periodic pattern, which is consistent across runs. Upon further inspection, we found that the corresponding blocks aggregate data associated to CPU cycles, and as such this pattern indicates oscillating CPU frequencies on all nodes, induced by Quicksilver's code mix. The AMG application shown on the right side of Figure~\ref{method:csflowchart} exhibits yet a different behavior: in the upper half of the real components a gradient in the block values can be observed, which is associated with increasing memory usage over the run. Similarly to Kripke, AMG shows a distinct iterative behavior as well.

The structure of the CS signatures is consistent in all figures and matches our description in Section~\ref{section:algorithm}: the upper section includes blocks that are descriptive of system behavior, while the middle part groups those carrying little information content, with metrics being almost constant or unaffected by the system's status. Finally, the blocks in the bottom part are again descriptive of system behavior, but their values are inversely proportional to those of the blocks at the top. Like in a periodic signal, CS signatures are able to highlight periodic behaviors only where their period $p > 2 \cdot w^l$, in accordance with the sampling rate of the original data. Unlike the baseline methods, the CS method produces understandable data representations that can be used as a launch platform for additional low-level performance analyses, satisfying the \emph{Visualizability} requirement. 

\subsection{Portability across Architectures}
\label{section:portability}
Here we demonstrate the generality of the signatures computed with the CS method. To this end, we performed a test using the Cross-Architecture segment of HPC-ODA, replicating the classification task of the Application one. Here, the BEAST testbed, \sng and \cmuc nodes used to acquire data have different architectures and amounts of sensors: the testbed node is equipped with two 64-core AMD Epyc Rome 7742 CPUs and has 39 sensors, while the \sng node uses two 24-core Intel Skylake Xeon Platinum 8174 CPUs and has 52 sensors. Finally, the \cmuc node uses a 64-core Intel Xeon Phi 7210-F many-core CPU and has 46 sensors. Using a single node per hardware architecture allows us to focus on their peculiarities and the impact they have on monitoring and machine learning models, as opposed to network and application behavior in a distributed context; as a consequence, the underlying applications in the segment are executed in shared-memory OpenMP fashion. We process this heterogeneous data as in the following:

\begin{enumerate}
    \item The CS method is applied to the data of each of the 3 nodes independently, generating 20-block signatures.
    \item We merge the three resulting datasets in a single one.
    \item We perform 5-fold cross-validation, classifying running applications with no knowledge of the architecture.
\end{enumerate}

\begin{figure}[t]
\centering
\captionsetup[subfigure]{}
\subfloat[BEAST testbed node.]{
\includegraphics[width=0.23\textwidth,trim={10 0 120 0}, clip=true]{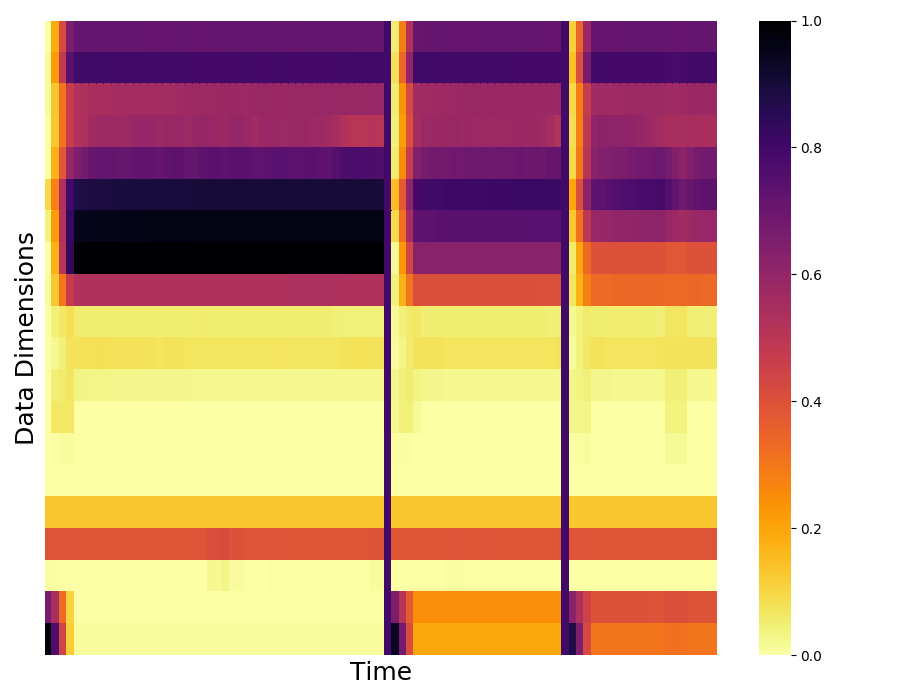}
\includegraphics[width=0.23\textwidth,trim={6 0 120 0}, clip=true]{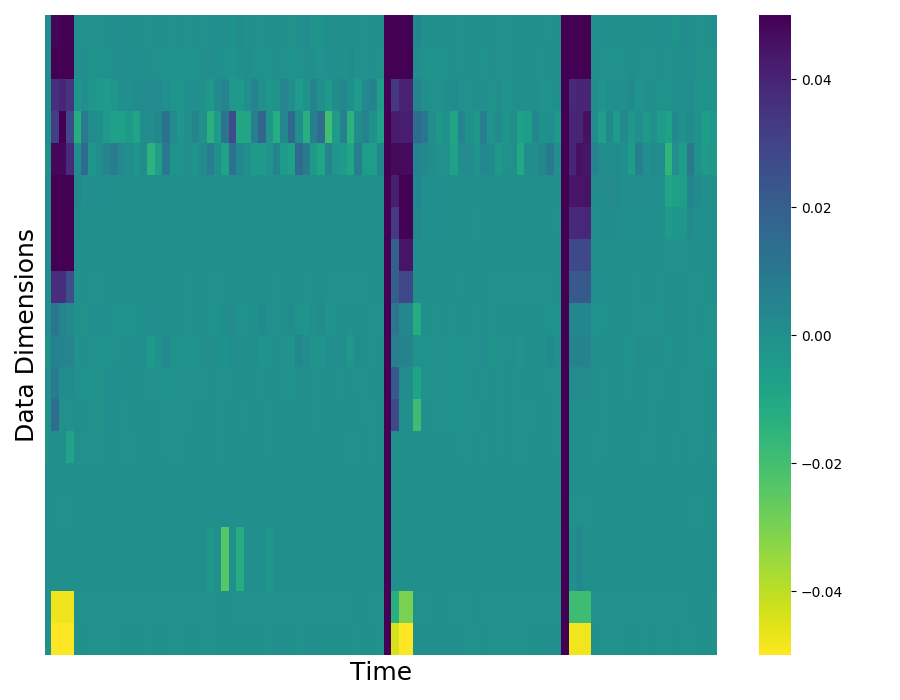}
  }
\\
\subfloat[\sng node.]{
\includegraphics[width=0.23\textwidth,trim={10 0 120 0}, clip=true]{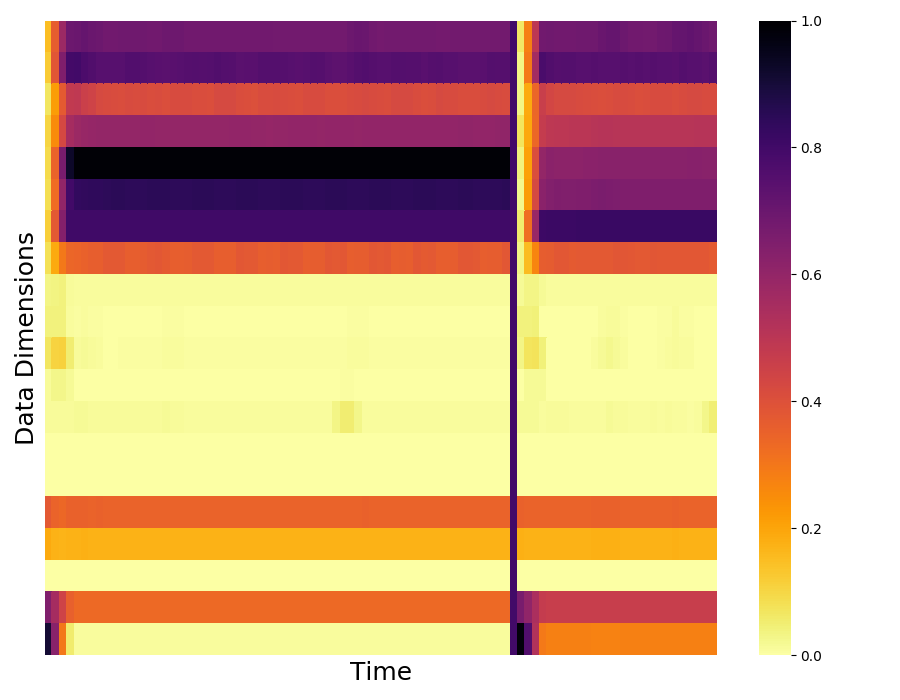}
\includegraphics[width=0.23\textwidth,trim={6 0 120 0}, clip=true]{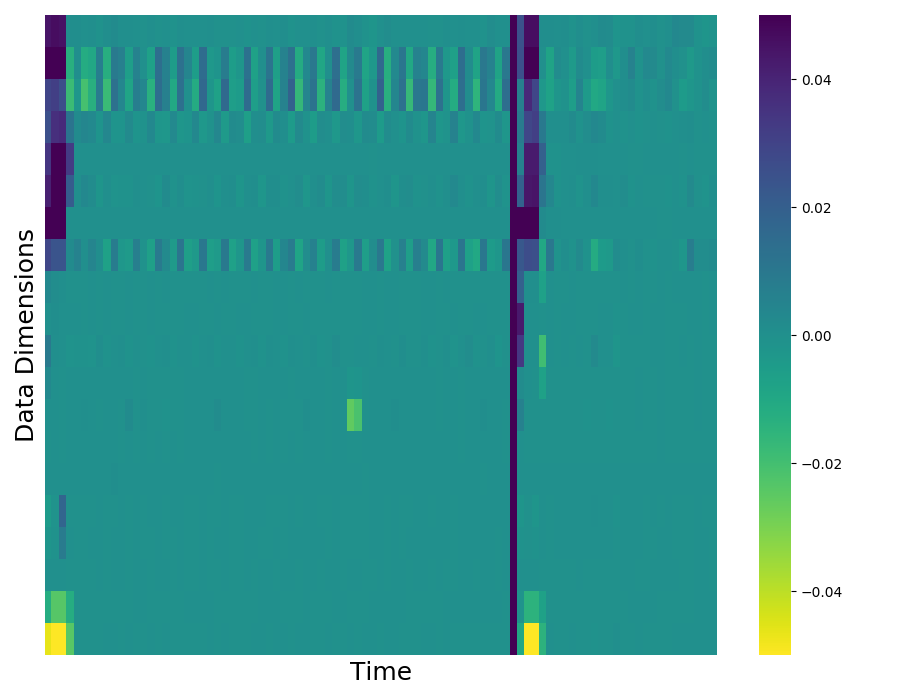}
  }
\\ 
\subfloat[\cmuc node.]{
\includegraphics[width=0.23\textwidth,trim={10 0 120 0}, clip=true]{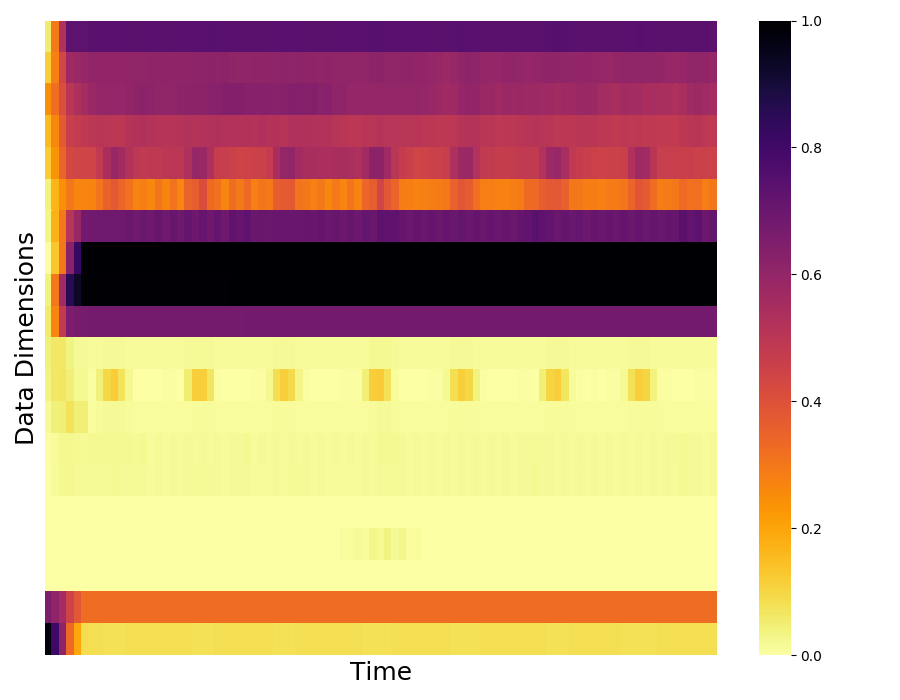}
\includegraphics[width=0.23\textwidth,trim={6 0 120 0}, clip=true]{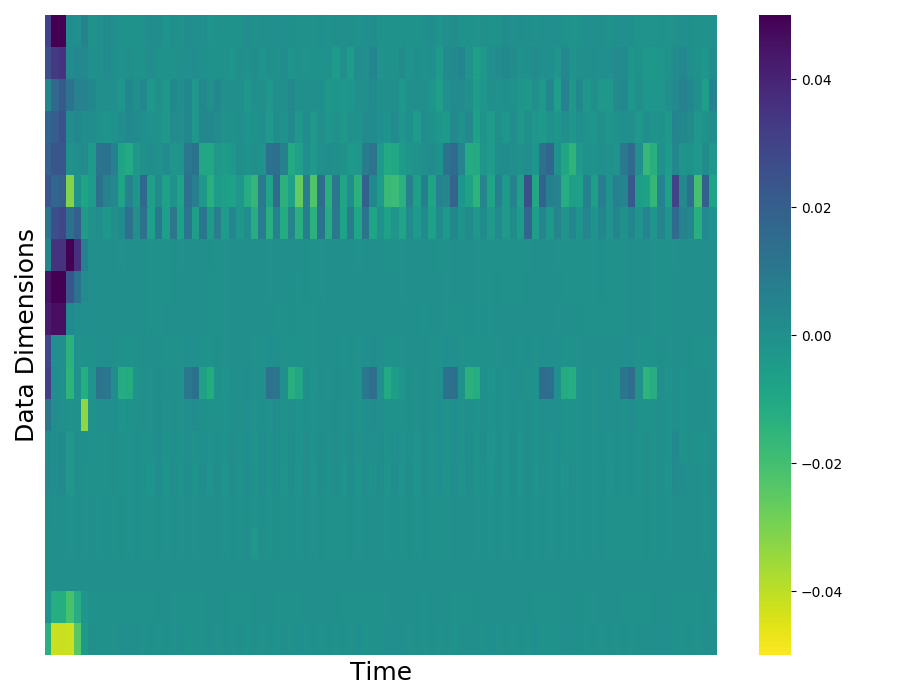}
  }
\caption{Real (left) and imaginary (right) components of the signature heatmaps for the LAMMPS application on three different compute node types, using the CS method with 20 blocks. Solid vertical lines indicate the end of a run.}
\label{results:visualizations2}
\end{figure}

This experiment led to an average F1-score of 0.995 using a random forest model, a result that is equivalent to the one presented for the Application segment with no significant degradation in performance. When using a multi-layer perceptron classifier, the F1-score amounts to 0.992. This confirms that the CS method produces generic signatures and enables portability of models across systems, as well as robustness against changes in the sets of available sensors. In Figure~\ref{results:visualizations2} we show the signature heatmaps associated with the LAMMPS application, executed under the same configuration on each of the 3 compute nodes. The format of the signatures changes slightly across the architectures, and there are obvious execution time differences due to the varying performance of the nodes. However, the same performance patterns can be recognized in all cases, both in terms of periodicity in the imaginary parts and distribution of the metrics in the real ones. This is confirmed upon further inspection of the sensors' ordering after applying the CS method, which was consistent and similar even though the data of each compute node was treated separately. It should be noted that this experiment cannot be reproduced at all using the baseline methods in Section~\ref{section:baseline}, since they produce incompatible signatures with different lengths depending on the specific compute node: instead, the CS method allowed us to train a single model to recognize applications under different architectures and monitoring configurations, satisfying the \emph{Portability} requirement.

\section{Conclusions}
\label{section:conclusions}

In this work we presented the Correlation-wise Smoothing (CS) method for extracting knowledge from multi-dimensional time-series data. The CS method is tailored for the specific needs of data center monitoring and ODA, where costly data mining methods are often unusable due to strict operational requirements. It uses an effective algorithm that re-arranges data dimensions in order to group metrics that are correlated to one another. This produces image-like representations that can be scaled, manipulated and visualized at will and with very little computational cost. We evaluate the CS method and demonstrate its properties using the HPC-ODA dataset collection, which we release publicly with this work. The CS method shows several advantages over current state-of-the-art methods: it retains the same machine learning performance in up to one order of magnitude less time; it produces signatures that are up to one order of magnitude smaller; the signatures can be visualized effectively; finally, they are general enough to enable comparability and portability of models across systems, while showing predictable scaling properties.

As future work, we plan to further characterize the properties and performance of the CS method in scenarios that we could not explore in this work, such as the following:

\begin{itemize}
    \item Testing the CS method's effectiveness when applied to accelerator sensor data (e.g., GPUs) or to data coming from an entire HPC system running multiple user jobs. 
    \item Deploying a CS-based ODA control loop on a production HPC system, using a dedicated online implementation.
    \item Characterizing the overhead and scalability of the CS method in a production HPC environment.
    \item Exploring use cases outside of the data center domain.
\end{itemize}

\vspace{2mm}
\textit{Acknowledgements.} This research activity has received funding from the DEEP-EST project under the EU H2020-FETHPC-01-2016 Programme grant agreement n. 754304.

\bibliographystyle{IEEEtran}
\bibliography{main}

\end{document}